\documentclass[%
 aps,
 prl,
 amsmath,amssymb,
 reprint,
 superscriptaddress,
 longbibliography,
]{revtex4-1}

\usepackage{graphicx}
\usepackage{float}
\usepackage{dcolumn}
\usepackage{bm}
\usepackage{natbib,hyperref}
\hypersetup{colorlinks=true, allcolors=blue}
\usepackage{todonotes}
\usepackage{mhchem}

\begin{document}

\title{Transient dual-energy lasing in a semiconductor microcavity} 

\author{Feng-Kuo Hsu}
\author{Wei Xie}
\affiliation{Department of Physics and Astronomy, Michigan State University, East Lansing, MI 48824, USA}
\author{Yi-Shan Lee}
\author{Sheng-Di Lin}
\affiliation{Department of Electronics Engineering, National Chiao Tung University, Hsinchu 30010, Taiwan}
\author{Chih-Wei Lai}
\email{cwlai@msu.edu}
\affiliation{Department of Physics and Astronomy, Michigan State University, East Lansing, MI 48824, USA}

\date{\today}

\begin{abstract}
We demonstrate sequential lasing at two well-separated energies in a highly photoexcited planar microcavity at room temperature. Two spatially overlapped lasing states with distinct polarization properties appear at energies more than 5 meV apart. Under a circularly polarized nonresonant 2 ps pulse excitation, a sub-10-ps transient circularly polarized high-energy (HE) state emerges within 10 ps after the pulse excitation. This HE state is followed by a pulsed state that lasts for 20--50 ps at a low energy (LE) state. The HE state is highly circularly polarized as a result of a spin-preserving stimulated process, while the LE state shows a significantly reduced circular polarization because of a diminishing spin imbalance.
\end{abstract}

\pacs{Valid PACS appear here}

\maketitle

The spectral characteristics of a conventional semiconductor laser are typically fixated to static composition structures \cite{koch1995,chow1999,iga2003,michalzik2013a}. As a result, the lasing energy is generally single-valued, with a sub-1-meV linewidth. Nevertheless, a simultaneous two-state lasing effect with well-separated lasing wavelengths has been demonstrated in quantum dot lasers \cite{markus2003,markus2006,viktorov2005} and nanocrystal lasers \cite{chan2004} when the intraband carrier relaxation between the ground and excited states quasi-0D nanoscale gain medium is finite. However, the lasing energies in these quantum-dot lasers are still fixated to a static cavity structure. Another essential characteristic of a semiconductor laser is polarization. In edge-emitting semiconductor lasers, laser radiation is typically linearly polarized as determined by the polarization-dependent reflectivity of the cavity. By contrast, in vertical-cavity surface-emitting lasers (VCSELs), the lasing polarization is inherently unstable and is affected by gain or crystalline anisotropy \cite{chang-hasnain1991,panajotov2013} of the gain medium or cavity. VCSELs typically give laser radiation linearly polarized either along the [110] or [110] crystallographic direction, but can exhibit complex polarization switching \cite{san-miguel1995,ostermann2013,panajotov2013} and bistability \cite{tang1987,wieczorek2005} effects depending on the cavity structure and carrier density.

In this study, we report optically controlled dual-energy (two-state) lasing in a highly photoexcited planar Fabry-P\'erot semiconductor microcavity in which the gain media, In$_{0.15}$Ga$_{0.85}$As quantum wells (QWs), are positioned at the antinodes of the cavity light field \cite{hsu2013a}. The sample is \emph{nonresonantly} photoexcited by circularly polarized 2 ps laser pulses at $E_p$ = 1.58 eV, which is about 250 meV above the QW bandgap ($E'_g \approx $ 1.33 eV; the $e1$--$hh1$ transition between the first quantized electron and heavy-hole levels) and 170 meV above the bare cavity resonance ($E_c \approx$ 1.410 eV). Above a critical photoexcited density ($n_c$), near-unity circularly-polarized $\sim$10 ps pulsed radiation appears within 10--20 ps after the pulse excitation. The spin-polarized lasing is attributed to the spin-dependent stimulation of cavity-induced correlated electron-hole (\emph{e-h}) pairs formed near the Fermi edge of the high-density \emph{e-h} plasmas \cite{hsu2015a}. The lasing energies of the samples studied range from $E_\mu =$ 1.40 eV to 1.42 eV, which are near the transition energy between the second quantized energy levels in QWs ($E_g'' \approx 1.41$ eV; the $e2$--$hh2$ transition), but not locked to the bare cavity resonance $E_c$. 
 
\begin{figure}[htbp!]
\includegraphics[width= 0.4\textwidth]{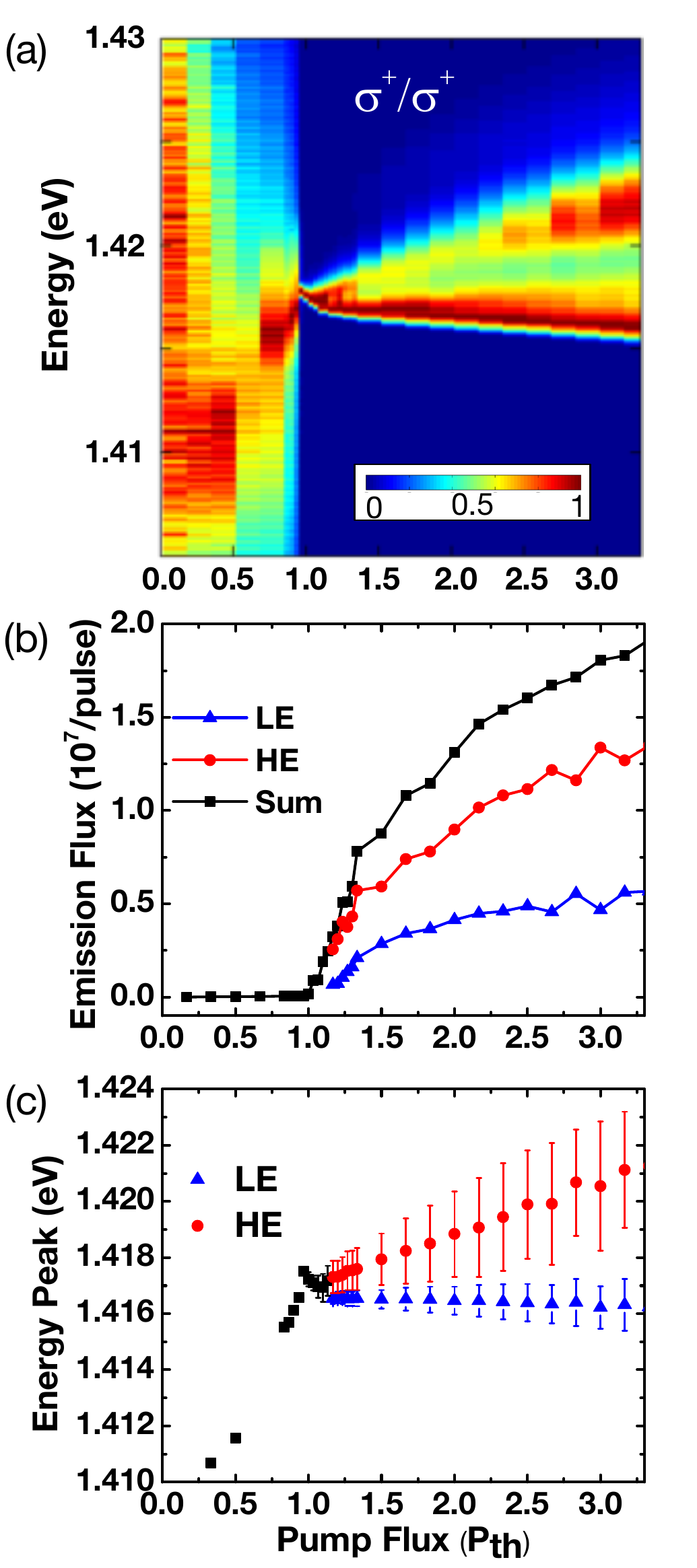}
\caption{ \textbf{Spectral characteristics of dual-energy (two-state) lasing.} (a) Normalized time-integrated spectra of the co-circular ($\sigma^{+}/\sigma^{+}$) emission component for $|k_\parallel|<3$ $\mu$m$^{-1}$. (b) Emission flux of the HE state (solid red circles), the LE state (solid blue triangles), and the sum (solid black squares). (c) Peak energies (solid shapes) and linewidths (error bars) of the HE state (solid red circles) and LE state (solid blue triangles). The peak energies below the threshold are represented by the solid black squares. The emission fluxes, peak energies, and linewidths are determined by fitting of the spectra with multiple-Gaussian functions. The photoexcited density at the threshold pump flux ($P_{th}$) is $n_c  \approx 2 \times 10^{12}$ cm$^{-2}$ per quantum well \emph{per pulse}.
}\label{fig:spec_analysis}
\end{figure}

In spatially localized areas \footnote{All samples display a spin-polarized state; however, the localized LE state typically occurs more frequently in samples in which $E_g''$ exceeds $E_c$. In this study, we focus on a sample with a lasing energy of 1.417 eV at the threshold.}, an additional low-energy (LE) lasing state with distinct polarization properties appears at well-separated energies of up to 5 meV apart when the photoexcited density increases to a bifurcation density ($n_b \approx 1.1 \ n_c$) (Fig.~\ref{fig:spec_analysis}). The two transient lasing states with distinct spectral and polarization characteristics emerge sequentially in time. Temporally, radiation from the high-energy (HE) spin-polarized state commences within 10 ps after pulse excitation and lasts for $\sim$10 ps. The LE state follows the HE state and lasts for 20--50 ps. Spectrally, slightly above the threshold, the HE state has a time-integrated linewidth $\lesssim$ 3 meV, whereas the LE state has a linewidth of less than 1 meV. With increasing photoexcited density, the HE state blueshifts about 5 meV, while the LE state redshifts less than 1 meV. As a result, the energy separation between these two states depends on the photoexcited density. Moreover, the two lasing states have distinct polarization properties that are largely determined by the spin imbalance in the \emph{e-h} plasmas. Slightly above the lasing threshold, the HE state gives fully circularly polarized radiation as a result of a spin-preserving stimulated process. With an increasing photoexcited density, the HE state becomes a spinor-like state and gives pulsed radiation of positive ($+$) and negative ($-$) helicity with a spin-splitting of about 1 meV. By contrast, the LE state appears 20--50 ps after the pulse excitation and exhibits a vanishing degree of circular polarization because of the diminishing spin imbalance. 
 
\begin{figure}[htb]
\includegraphics[width= 0.48 \textwidth]{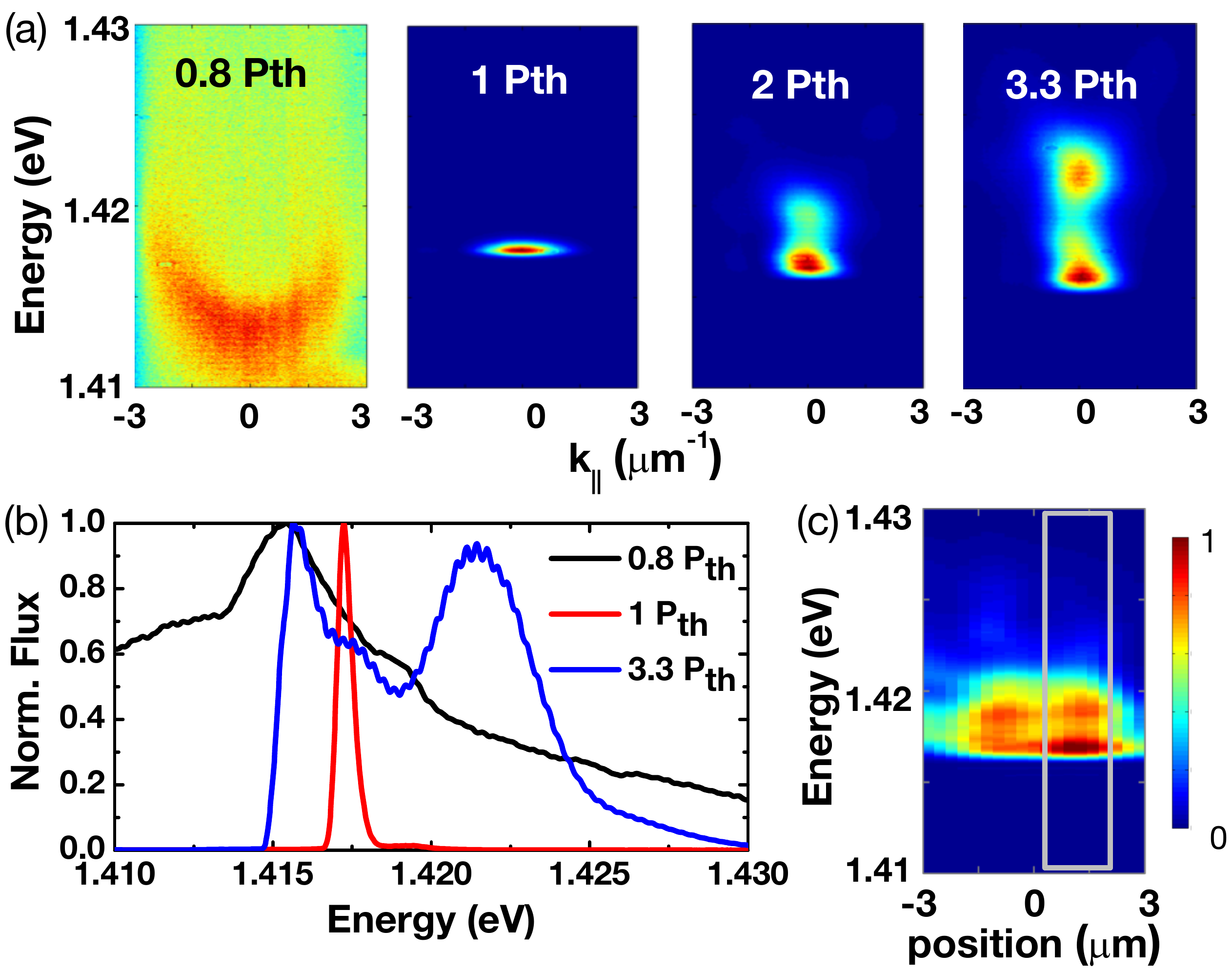}
\caption{\textbf{Two-state lasing in a microcavity.} (a) Angularly resolved (k-space) imaging spectra of the co-circular ($\sigma^{+}/\sigma^{+}$) emission component at $P$ = 0.8, 1.0, 2.0, and 3.3 $P_{th}$. (b) Time-integrated spectra at 0.8 (solid black line), 1.0 (solid red line) and 3.3 $P_{th}$ (solid blue line). (c) Real-space (r-space) imaging spectra at 2.0 $P_{th}$. The k-space spectra shown in (a--b) are measured through a pinhole, as represented by the gray box. 
}\label{fig:ek_spec}
\end{figure}

\paragraph*{Methods.}
The undoped microcavity consists of a $\lambda$ cavity sandwiched within two Bragg mirrors with alternating AlAs/GaAs layers with thickness of $\lambda/4$. The active layer includes three sets of three In$_{0.15}$Ga$_{0.85}$As/GaAs (6-nm/12-nm) multiple quantum wells (QWs) embedded at the antinodes of the cavity light field within a $\lambda$ GaAs cavity. The QW bandgap ($E_g'$) is tuned through a rapid thermal annealing process at 1010$^\circ$C--1090$^\circ$C for 5--10 s, in which $E_g'$ blueshifts up to about 20 meV as a result of the diffusion of gallium ions into the QWs.  A Fourier transform optical system is used for angle-resolved (k-space) and space-resolved (r-space) luminescence spectroscopy and polarimetry. The polarizations of the excitation and emission are controlled and analysed with liquid crystal devices. A circularly polarized light (pump or luminescence) with angular momentum $\pm\hbar$ along the pump laser wavevector $\hat{k} \parallel \hat{z}$ is defined as $\sigma^{\pm}$. Time-resolved luminescence and polarimetry are performed with a combination of an imaging spectrometer and ps streak camera system. Further descriptions of the composition structure and the experimental setups are provided in Ref. \cite{hsu2013a}. 

\paragraph*{Results.}
We first analyse the density-dependent energy shifts and linewidths. Fig.~\ref{fig:spec_analysis}a shows the co-circular $\sigma^{+}/\sigma^{+}$ luminescence spectra [$S^+(E)$] at $k_{\parallel}\approx 0$ as a function of pump flux. Above the bifurcation pump flux $P_{b} \equiv 1.1 \ P_{th}$, radiation bifurcates into a doublet (\emph{two-state lasing}) with a dominant HE state. The emission flux of the LE state reaches a plateau near above $1.5 \, P_{th}$, whereas the emission flux of the HE state continue to increase linearly with the pump flux. Spectrally, the HE and LE states display distinct energy shifts with increasing photoexcited density. Fig.~\ref{fig:spec_analysis}c shows the peak energies of the two states with increasing pump flux. Below $P_b$, the emission is single-peaked and blueshifts by 6 meV when the pump flux increases from 0.5 $P_{th}$ to $\sim1.1 \ P_{th}$. When the pump flux increases gradually from 1.0 $P_{th}$ to 3.5 $P_{th}$, the HE state blueshifts linearly by 5 meV, whereas the LE state redshifts by less than 0.5 meV. This result indicates that the energy difference between the HE and LE states increases with increasing photoexcited density. Above $P_b$, the linewidth of the HE state increases from $\sim$0.3 meV at $P_{th}$ to 3 meV at 4 $P_{th}$. On the other hand, the LE state remains spectrally narrow with a linewidth of 1 meV or less.

Next, we study the spectral characteristics in k-space and r-space. Fig.~\ref{fig:ek_spec}a shows the k-space imaging spectra of the co-circular emission component for selected pump fluxes. A nearly parabolic energy versus in-plane momentum ($E$ vs. $k_{\parallel}$) dispersion curve emerge slightly below the threshold ($P \approx 0.8 \ P_{th}$). At the threshold (pump flux $P = P_{th}$), the radiation becomes directional (angular-spread $\Delta k_\parallel \leq 1 \mu m^{-1}$) and spectrally narrow (linewidth $\Delta E \leq $ 0.3 meV) (Fig.~\ref{fig:ek_spec}a--b). The time-integrated emission spectra appear to bifurcate into a doublet when the pump flux is increased above 2 $P_{th}$. The two lasing states are spatially overlapped in-plane, as evidenced in the r-space imaging spectra (Fig.~\ref{fig:ek_spec}c).

\begin{figure}[htb!]
\includegraphics[width=0.35\textwidth]{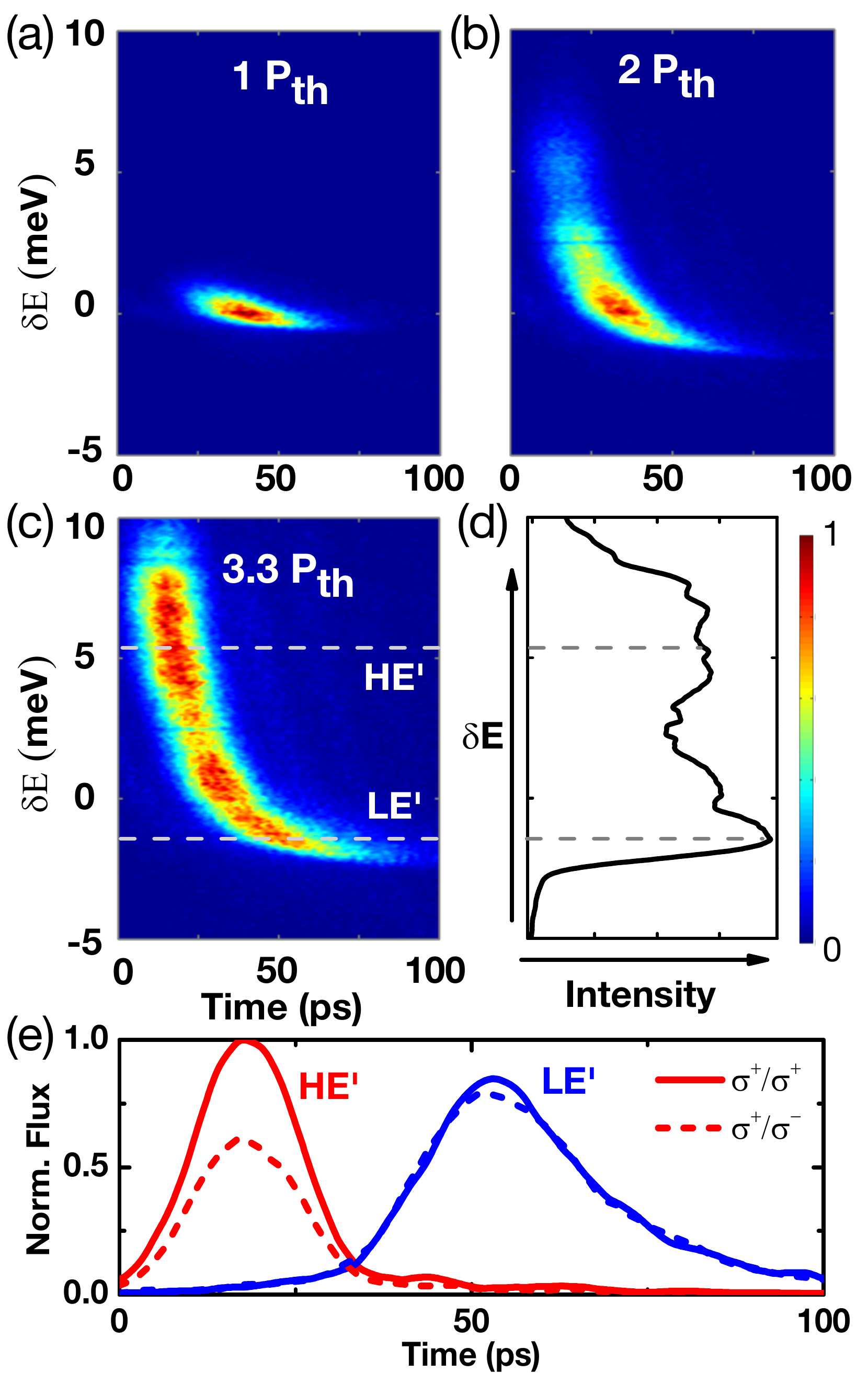}
\caption{\textbf{Dynamics.} (a--c) Time-dependent spectra of the co-circular component at $k_\parallel \approx 0$ under $P =$ 1.0, 2.0, and 3.3 $P_{th}$. The energy $\delta E$ is measured with respect to 1.415 eV, the lasing energy at the threshold. (d) The time-integrated spectrum obtained from (c). (e) Polarized time-dependent luminescence of the HE state at $\delta E =$ 5 meV [$HE'$ as indicated in (c)] (solid and dashed red lines) and the LE state $\delta E =$ $-5$ meV [$LE'$ as indicated in (c)] (solid and dashed blue lines) for $P =$ 3.3 $P_{th}$.}
\label{fig:dynamics}
\end{figure}

The spectral doublets that appears in time-integrated spectroscopy measurements are in fact temporally separated, as shown in the time-dependent polarized luminescence spectra at $k_{\parallel} \approx 0$ (Fig.~\ref{fig:dynamics}). When two-state lasing occurs, luminescence from the high-energy HE state appears within 10 ps after pulse excitation and decays with a time constant $\tau_d' \sim$10 ps. After the HE state diminishes, the low-energy LE state appears and then decays with a time constant $\tau_d'' \sim$30--50 ps. 

\begin{figure}
\centering
\includegraphics[width=0.35\textwidth]{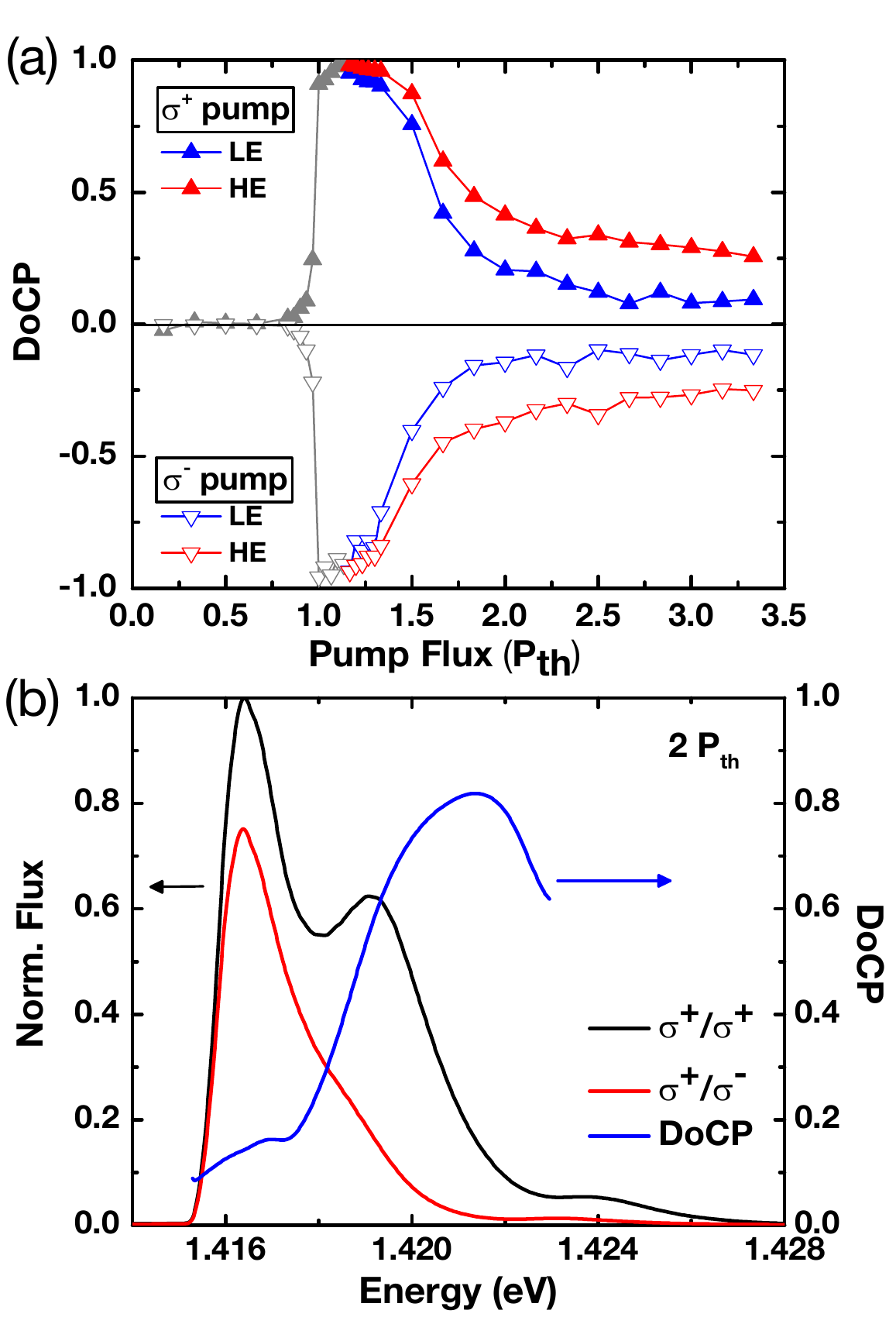}
\caption{
\textbf{Polarization Properties.} (a) Stationary degree of circular polarization ($DoCP = \bar{\rho}_c$) of the HE state (solid and open red triangles) and LE state (solid and open blue triangles) under circularly polarized $\sigma^+$ pump (upper part, solid triangles) or $\sigma^-$ pump (lower part, open triangles). $DoCP = \bar{\rho}_c = (A^+-A^-)/(A^++A^-)$, where $A^\pm$ refers to the areas of co-circular ($A^+$) and cross-circular ($A^-$) components of the HE and LE spectral peaks and are obtained by fitting of the time-integrated spectra (Fig.~\ref{fig:spec_analysis}) with multiple-Gaussian functions. The $\bar{\rho}_c$ of the HE states increases from zero to near unity for $P_{th} < P < 1.2 P_{th}$ and decreases gradually to 0.25. The $\bar{\rho}_c$ of the LE states is relatively high ($>$0.8) initially, but it decrease rapidly to less than 0.1 when the pump flux is increased above 2.0 $P_{th}$. (b) The time-integrated polarized spectra $S^\pm(E)$ of the co-circular ($\sigma^+/\sigma^+ \rightarrow S^+(E)$, black curve) and the cross-circular ($\sigma^+/\sigma^- \rightarrow S^-(E)$, red curve) components at $P = P_{th}$. The energy-dependent $DoCP(E) = \bar{\rho}_c(E) = [S^+(E)-S^-(E)]/[S^+(E)+S^-(E)]$ (blue curve) shows a maximal $\bar{\rho}_c(E) \approx$ 0.8, which is significantly larger than the spectrally averaged $\bar{\rho}_c$ shown in (a). 
}
\label{fig:pol}
\end{figure}

In Fig.~\ref{fig:pol}, we study the polarization properties of the two lasing states. Fig.~\ref{fig:pol}a shows the $\bar{\rho}_c$ of the HE and LE states as a function of pump flux. The $\bar{\rho}_c$ of the HE state rises from nearly zero to unity at threshold, remains above 0.9 for $P_{th}< P < 1.2 \ P_{th}$, and decreases gradually to 0.25 at 4 $P_{th}$. The LE state appears with a sizable $\bar{\rho}_c$ slightly above 1.1 $P_{th}$, and shows rapidly diminishing $\bar{\rho}_c$ for $P \gtrsim 1.7 P_{th}$. Moreover, an apparent spin-dependent energy splitting of $\sim$1 meV between the co-circular and cross-circular components of the HE state appears, as demonstrated by the selected polarized emission spectra at 2.0 $P_{th}$ shown in Fig~\ref{fig:pol}b (see also Ref.~\cite{hsu2015a}). The spectrally resolved $DoCP(E) = \bar{\rho}_c(E)$ also shows the energy-dependent $\bar{\rho}_c$, in which the HE state reaches a maximal $\bar{\rho}_c(E) \approx$ 0.8, even at a high photoexcited density. Such a high $\bar{\rho}_c(E)$ occurs despite the sub-10-ps spin relaxation times of electrons and holes in InGaAs/GaAs QWs, a result that is indicative of sub-10-ps carrier cooling and spin-dependent stimulated processes. The rapid cooling results in a sizable spin imbalance in the cavity-induced correlated \emph{e-h} pairs formed near the Fermi edge of the high-density plasmas, whereas the spin-dependent stimulation of these \emph{e-h} pairs amplifies such optical spin polarization in the presence of non-radiative loss \cite{hsu2015a}. By contrast, the LE state has a vanishing $\bar{\rho}_c$ because of the lack of spin imbalance at a time delay of 50 ps or more after the optical injection of the spin-polarized carriers.

\paragraph*{Discussion.}
In this study, the HE state is dominated by a macroscopic population of ``hot" \emph{e-h} pairs near the Ferm edge via the spin-dependent stimulation. The LE state forms in a confined area of a lateral dimension of $\sim$2--5$\mu$m, which is defined by natural crystalline in-plane inhomogeneities (Fig.~\ref{fig:ek_spec}). After the radiative recombination of these ``hot" \emph{e-h} pairs, the remaining ``cold" \emph{e-h} pairs form the LE state that gives spectrally narrow and directional radiation. Therefore, the formation of the LE state hinges on the population of the remaining ``cold" \emph{e-h} pairs, which typically becomes significant in samples in which the chemical potential $\mu$ can exceed the bare cavity resonance significantly (i.e., $\mu \sim E_g'' \gtrsim E_c$). 

We note that spectral functions resembling the two-state lasing spectra presented here have been calculated for \emph{e-h} systems in the presence of a cavity by Littlewood et al. \cite{eastham2001} and Ogawa et al. \cite{kamide2011,kamide2012} . Excitations in exciton condensates \cite{zhu1996,littlewood1996,eastham2001,wouters2007} or BCS-like \emph{e-h} states \cite{keeling2005,kamide2010,kamide2011,kamide2012,yamaguchi2012,yamaguchi2013} can give rise to spectral multiplets. Analogous frequency shifts and doubling of the optical excitations have been reported in a Bose gas \cite{jin1997,pethick2001,oktel1999}, in which condensate and thermal components coexist. In general, the doublet in the absorption spectra for normal and condensate components in a cold Bose gas is fundamentally related to the two-fluid model of superfluid helium \cite{donnelly2009} and the two-component superconductivity model of high $T_c$ superconductors \cite{baryam1990}. 

To clarify the mechanisms of the two-state lasing effect in the samples studied here, conducting further temporally and spatially resolved spectroscopies to characterize the energy distribution and relaxation of nonradiative \emph{e-h} carriers is necessary. For example, one can use pump-probe spectroscopy to show any potential resonance or gain--absorption gap near the Fermi edge of the high-density \emph{e-h} plasma coupled to the cavity light field.

\begin{acknowledgements}
We thank Mark Dykman, Brage Golding, Leonid S. Levitov, Peter B. Littlewood, John A. McGuire, and Carlo Piermarocchi for the discussions. This work was supported by NSF grant DMR-09055944 as well as a start-up funding and the Cowen endowment at Michigan State University.
\end{acknowledgements}

\bibliography{/Users/cwlai/Dropbox/Bib/lai_lib}

\end{document}